\documentclass[aip, amsmath,amssymb,english]{revtex4}
\usepackage[latin9]{inputenc}
\usepackage{amsthm}
\usepackage{amsmath}
\usepackage{amssymb}
\usepackage{amstext}
\usepackage{esint}

\makeatletter

\theoremstyle{plain}
\newtheorem{definition}{Definition}
\newtheorem{proposition}{Proposition}
\newtheorem*{thm*}{Theorem}
\newcommand{\be}{\begin{equation}}
\newcommand{\nd}{\noindent}
\newcommand{\ee}{\end{equation}}
\newcommand{\ben}{\begin{eqnarray}}
\newcommand{\een}{\end{eqnarray}}

\makeatother

\usepackage{babel}

\begin{document}

\title{On a conjecture about Dirac's delta representation using q-exponentials}

\author{A. Chevreuil}
\affiliation{E.S.I.E.E., Universit\'{e} de Marne la Vall\'{e}e, Marne la Vall\'{e}e, France}
\author{A. Plastino}
\affiliation{C.C.T.-Conicet, National University La Plata,
C.C. 727, 1900 La Plata, Argentina}
\author{C. Vignat}
\affiliation{L.S.S., Supelec, France}

\begin{abstract}

A new representation of Dirac's delta-distribution, based on the
so-called q-exponentials, has been recently conjectured. We prove
here that this conjecture is indeed valid.

\end{abstract}

 \maketitle

\section{Introduction}

Tsallis and Jauregui have recently conjectured  \cite{tsallis} a
representation of the Dirac delta distribution, which they call
$\delta_q(x)$, based on $q-$exponential functions.
However, they could not prove their conjecture and used numerical experiments that suggest its
validity.  In this note, we provide a rigourous mathematical approach to this problem and prove
their conjecture by recourse to the notion of superstatistics.

\section{$q-$ exponentials and superstatistics}

\nd Statistical Mechanics' most notorious and renowned probability
distribution is that deduced by Gibbs for the canonical
ensemble \cite{reif, pathria}, usually referred to as the
Boltzmann-Gibbs equilibrium distribution

\be \label{gibbs}  p_G(i) =\frac{\exp{(-\beta E_i)}}{Z_{BG}}, \ee
with $E_i$ the  energy of the microstate labeled by $i$,
$\beta=1/k_B T$ the inverse temperature, $k_B$ Boltzmann's
constant, and $Z_{BG}$ the partition function. The exponential
term $F_{BG}=\exp{(-\beta E)}$ is called the Boltzmann-Gibbs
factor. Recently Beck and Cohen \cite{super} have advanced a
generalization, called superstatistics,  of this BG factor,
assuming that the inverse temperature $\beta$ is a stochastic
variable. The generalized statistical factor $F_{GS}$ is thus
obtained as the multiplicative convolution \be \label{bc}
F_{GS}=\int_0^\infty\, \frac{d\beta}{\beta}\,f(\beta)\,
\exp{(-\beta E)},\ee where $f(\beta)$ is the density probability
of the inverse temperature.


\nd As stated above, $\beta$ is the inverse temperature, but the
integration variable may also be any convenient intensive
parameter.  Superstatistics, meaning ``superposition of
statistics", takes into account fluctuations of such intensive
parameters.

\nd Beck and Cohen also show that if $f(\beta)$ is a Gamma
distribution, nonextensive thermostatistics is obtained,
which is of interest because this thermostatistics is today a very active field,
with applications to several scientific disciplines \cite{gellmann,lissia,fromgibbs}.
In working in a nonextensive framework,  one has to deal with power-law distributions,
which are certainly ubiquitous in physics (critical phenomena are
just  a conspicuous example \cite{goldenfeld}).
Indeed, it is well known that  power-law distributions arise when
maximizing Tsallis'  information measure
\be
H_{q}\left(  f\right)  =\frac{1}{1-q}\left(  1-\int_{-\infty}^{+\infty}
f(x)^{q} dx\right), \label{dino}
\ee
subject to appropriate constraints, where $q\ne1$ is a real positive parameter called the nonextensivity index.
More precisely, in the case of the canonical distribution, there is
only one constraint, the energy $E$, i.e. $\langle X^{2}\rangle =
E>0,$ and  the equilibrium
 canonical distribution writes in the case $q>1$

\[
f_{q}(x)=\frac{1}{Z_{q}}\left(1-(1-q)\beta_q x^{2}\right)^{\frac{1}{1-q}},
\]
where $\beta_q$ and
$Z_q$ stand for the nonextensive counterparts of $\beta$ and $Z_{BG}$
above. 

In the rest of this paper, we'll assume as in \cite{tsallis} that $1<q<2.$
\\
Let us choose the banch cut $\left] -\infty, -\frac{1}{1-q}\right[$ along the negative real  axis and  define the $q-$exponential function for $z\in\mathbb{C} \diagdown \left] -\infty, -\frac{1}{1-q}\right[$ as
\begin{equation}
\label{qexp}
e_{q}\left(z\right)=\left(1+ \left(1-q\right) z \right)^{\frac{1}{1-q}}.
\end{equation}
This allows us to rewrite the equilibrium distribution in the more natural way
\[
f_{q}\left(x\right)=\frac{1}{Z_{q}} e_{q}\left(-\beta x^{2} \right).
\]
It is a classical result that as $q\rightarrow1^{+},$   Tsallis
entropy reduces to
Shannon entropy
\be \label{shannon} H_{1}\left(  f\right)
=-\int_{-\infty}^{+\infty}f(x)\log f(x). \ee
Accordingly, the $q-$exponential function $e_{q}(x)$  converges to the usual exponential function $e^{x}$.



\section{Proof of Jauregui-Tsallis' conjecture }
\subsection{Definitions and Notations}
\nd Recall the  formula

\begin{equation}
\delta(t)=\frac{1}{2\pi}\int_{\mathbb{R}}e^{-\imath ut}du.\label{eq:dirac1}
\end{equation}
We intend to provide a generalization of this relation; namely, we
prove
 the following representation conjectured
by Tsallis \emph{et al.} assuming $1< q<2$:

\begin{equation}
\delta(t)=\frac{1}{c_{q}}\int_{\mathbb{R}}e_{q}\left(-\imath ut\right)du.\label{eq:dirac2}
\end{equation}
for some constant $c_{q}$. We begin by  recalling the mathematical
meaning of \eqref{eq:dirac1}.

\begin{definition}
A function $\varphi$ is called rapidly decreasing if $\varphi$ is $\mathcal{C}^{\infty}$
and if for all integers $k,\ell$
\[
\lim_{x\to\pm\infty}x^{k}\varphi^{(\ell)}(x)=0.
\]
\end{definition}
Let $\mathcal{S}$ be the set of the rapidly decreasing functions
on $\mathbb{R}$ and by $\mathcal{S}'$ the set of the continuous
linear functionals over $\mathcal{S}.$ For
$\varphi\in\mathcal{S},$ its Fourier transform
$\mathcal{F}(\varphi)$ is denoted by $\hat{\varphi}.$
\\

\nd We know from Rudin \cite[p. 184 theorem 7.4]{rudin} the
following
\begin{proposition}
\label{pro:We-know-from}
The Fourier transform $\mathcal{F}$ is a continuous linear mapping of  $\mathcal{S}$ into  $\mathcal{S}$.
\end{proposition}

\begin{definition}
Let now $f$ be a bounded measurable function \footnote{We could
extend the result to slowly increasing functions, but we do not
need this refinement here}. We let $T_{f}$ be the linear
continuous mapping:

\[
\forall\varphi\in\mathcal{S}\ \ \langle T_{f},\varphi\rangle=\int f(t)\varphi(t)dt
\]
\end{definition}

\subsection{Proofs}

\nd In order to prove the usual representation \eqref{eq:dirac1},
we simply have to show that for all $\varphi\in\mathcal{S}:$
\[
\int du \ \langle T_{e^{-\imath ut}},\varphi\rangle=2\pi\varphi(0).
\]
Of course $\langle T_{e^{-\imath ut}},\varphi\rangle=\hat{\varphi}(u)$. Hence the result.
\\

\nd We now turn to the proof of \eqref{eq:dirac2}. In this
respect, let us pick a $\varphi\in\mathcal{S}.$ We have
\begin{eqnarray*}
\langle T_{e_{q}(-\imath ut)},\varphi\rangle & = & \int e_{q}(-\imath ut)\varphi(t)dt\\
 & = & \int E_{W}e^{-\imath ut\left(q-1\right)W}\varphi(t)dt
\end{eqnarray*}
where we have used  the equality
\begin{equation}
\label{superstatistics}
e_{q}\left(-\imath ut\right)=E_{W}e^{-\imath ut\left(q-1\right)W}.
\end{equation}
Here
\[
E_{W}g(W)\triangleq\frac{1}{\Gamma(\frac{1}{q-1})}\int_{0}^{+\infty}g(w)e^{-w}w^{\frac{1}{q-1}-1}dw
\]
is the expectation of $g(W),$ where $W$ is a Gamma distributed random
variable with shape parameter $\frac{1}{q-1}$ and $g$ some function such that the above definition makes sense.

\nd We note that (\ref{superstatistics})  expresses the fundamental principle of the superstatistical theory.\\
On the other hand,
we have

\[
\frac{1}{\Gamma(\frac{1}{q-1})}\int\int e^{-w}w^{\frac{1}{q-1}-1}\vert\varphi(t)\vert dtdw\leq\int\vert\varphi(t)\vert dt.\]
 As obviously $\varphi$ is summable, we can apply the Fubini-Lebesgue theorem
and we obtain
\begin{eqnarray*}
\langle T_{e_{q}(-\imath ut)},\varphi\rangle & = & E_{W}\int e^{-\imath ut\left(q-1\right)W}\varphi(t)dt\\
 & = & E_{W}\hat{\varphi}(u\left(q-1\right)W)\\
 & = & \frac{1}{\Gamma(\frac{1}{q-1})}\int e^{-w}w^{\frac{1}{q-1}-1}\hat{\varphi}(u\left(q-1\right)w)dw
 \end{eqnarray*}

\nd Now, consider
\[
\int_{\mathbb{R}}\langle T_{e_{q}(-\imath ut)},\varphi\rangle du.
\]
As $q<2,$ we have, by the change of variable $u \mapsto v=u\left(q-1\right)w,$
\[
\int dw\int\vert e^{-w}w^{\frac{1}{q-1}-1}\hat{\varphi}(u\left(q-1\right)w)\vert du=\frac{\int\vert\hat{\varphi}(v)\vert dv}{q-1}\int e^{-w}w^{\frac{1}{q-1}-2}dw<\infty
\]
since, by Proposition \ref{pro:We-know-from},  $\hat{\varphi}\in\mathcal{S}$.
Thanks to the Fubini-Lebesgue theorem, we deduce that \begin{eqnarray*}
\int_{\mathbb{R}}\langle T_{e_{q}(-\imath ut)},\varphi\rangle du & = & \frac{1}{\Gamma(\frac{1}{q-1})}\frac{\int\hat{\varphi}(v)dv}{q-1}\int e^{-w}w^{\frac{1}{q-1}-2}dw\\
 & = & \frac{\Gamma(\frac{1}{q-1}-1)}{\left(q-1\right)\Gamma(\frac{1}{q-1})}2\pi\varphi(0)
 \end{eqnarray*}
We have proved the result with

\[
c_{q}=\frac{2\pi}{2-q}.\]

\section{Remarks on Jauregui-Tsallis' approach}
In their approach \cite{tsallis}, the authors chose an empirical
approach starting from the usual $q=1$ case: expressing the Dirac
delta as the limit
\[
\delta\left(x\right)=\frac{1}{c_{1}}\lim_{L\to+\infty}\int_{-L}^{+L}e^{-ikx}dk=\frac{2}{c_{1}}\lim_{L\to+\infty}\frac{\sin\left(Lx\right)}{x},
\]
the normalization constant $c_{1}$ can be obtained formally
as
\[c_{1}=2\lim_{L\to+\infty}\int_{-\infty}^{+\infty}\frac{\sin\left(Lx\right)}{x}dx=2\pi.\]

\nd The extension to $q-$exponentials reads
\[
\delta_{q}\left(x\right)=\frac{2}{\left(2-q\right)c_{q}}\lim_{L\to+\infty}\frac{\sin\left(\frac{2-q}{q-1}\arctan\left(\left(q-1\right)L\right)\right)}{x\left(1+\left(q-1\right)L^{2}x^{2}\right)^{\frac{2-q}{2\left(q-1\right)}}},
\]
so that the normalization constant $c_{q}$ can be obtained formally
as
\[
c_{q}=\frac{2}{2-q}\lim_{L\to+\infty}\int_{-\infty}^{+\infty}\frac{\sin\left(\frac{2-q}{q-1}\arctan\left(\left(q-1\right)L\right)\right)}{x\left(1+\left(q-1\right)L^{2}x^{2}\right)^{\frac{2-q}{2\left(q-1\right)}}}dx.
\]
This integral can be equivalently expressed, using the change of variable
$z=\tan\theta$ as\[
I_{q}\triangleq2\int_{0}^{\frac{\pi}{2}}\frac{\sin\left(\frac{2-q}{q-1}\theta\right)\left(\cos\theta\right)^{\frac{2-q}{q-1}-1}}{\sin\theta}d\theta.\]

\nd The authors then evaluate the integral $I_{q}$ for a finite number of rational values of the parameter $q\in\left]1,2\right[$ only, 
for which the symbolic computation software Maple gives the value $I_{q}=\frac{\pi}{2}.$ However, this approach can be circumvented given the fact that the integral $I_{q}$ can be found in \cite[3.638.3]{gradshteyn},
with the value
$I_{q}=\frac{\pi}{2}\,\,\forall q\in\left]1,2\right[$ so that empirically, $c_{q}=\frac{2\pi}{2-q}.$

\section{Conclusion}
\nd We have proved that the representation of the
Dirac delta distribution \eqref{eq:dirac2} using $q-$exponential
functions, as conjectured by Tsallis \emph{et al.}, is valid. In
particular, (i) we compute the exact normalization constant in the
representation of the Dirac delta and (ii) we explicit the set of
functions for which this distribution acts as the Dirac delta.




\begin{thebibliography}{1}

\bibitem{reif} Reif F 1965  {\it Statistical and thermal physics}
(NY: McGraw-Hill); Pathria R K 1993  {\it Statistical Mechanics}
(Exeter: Pergamon Press).


\bibitem{pathria}
 Gibbs J W  1948 {\it Elementary principles in statistical mechanics
in Collected Works} (New Haven: Yale University Press; R. B.
Lindsay R B and Margenau H 1957 {\it Foundations of physics\/}
 (NY: Dover).


\bibitem{super} C. Beck and E. G. D. Cohen, Physica A {\bf 322} (2003) 267.

\bibitem{gellmann} Gell-Mann M and Tsallis C, Eds. 2004
{\it Nonextensive Entropy: Interdisciplinary applications}
(Oxford: Oxford University Press) and references therein;
Plastino A R and  Plastino A 1994 Phys. Lett. A {\bf 193} 140.

\bibitem{lissia}  Kaniadakis G,  Lissia M, and  Rapisarda A., Eds., 2002
{\it Nonextensive statistical mechanics and physical
applications}, Physica A (Special) {\bf 305}, and references
therein.


\bibitem{fromgibbs} A.R.~ Plastino, A.~ Plastino, {\it Phys. Lett.
A} {\bf 193} (1994) 251.

\bibitem{tsallis}M.~Jauregui, C.~Tsallis, New representations of pi and Dirac delta using the nonextensive statistical mechanics q-exponential function, arXiv:1003.4967v3 [math-ph], to appear in Journal of Mathematical Physics

\bibitem{goldenfeld} N.~ Goldenfeld, {\it Lectures on phase transitions and the
renormalization group} (Addison-Wesley, NY, 1992).

\bibitem{rudin}W.~Rudin, {\it Functional Analysis}, Second Edition, McGraw-Hill, 1991

\bibitem{gradshteyn}I.S.~Gradshteyn and I.M.~Ryzhik, Table of Integrals, Series, and Products, Seventh Edition, Academic Press, 2007

\end{thebibliography}
\end{document}